# ELECTROCHEMICAL INSIGHTS INTO MANGANESE-COBALT DOPED α-Fe$_2$O$_3$ NANOMATERIAL FOR CHOLESTEROL DETECTION: A COMPARATIVE APPROACH

*Sushmitha S [a], Subhasmita Ray [b], Lavanya Rao [a], Mahesha P Nayak [a], Karel Carva [b], Badekai Ramachandra Bhat [a*]*

[a*] *Department of Chemistry, Catalysis and Materials Chemistry Laboratory, National Institute of Technology Karnataka, Surathkal, D.K., Karnataka 575 025, India.*

[b] *Department of Condensed Matter Physics, Faculty of Mathematics and Physics, Charles University, Ke Karlovu 3, Prague 12116, Czech Republic.*

*\*Corresponding Author: ram@nitk.edu.in*

**Abstract**

Herein, a self-assembled hierarchical structure of hematite (α-Fe$_2$O$_3$) was synthesized via a one-pot hydrothermal method. Subsequently, the nanomaterial was doped to get M$_x$Fe$_{2-x}$O$_3$ (M = Mn-Co; x = 0.01, 0.05, 0.1) at precise concentrations. The electrode was fabricated by coating the resulting nanocomposite onto a Nickel Foam (NF) substrate. The electrochemical characterization demonstrated the excellent performance of cobalt-doped α-Fe$_2$O$_3$, among which, Co$_{0.05}$Fe$_{0.95}$O$_3$ (CF5) exhibited superior performance, showing a two-fold increase in sensitivity of 1364.2 μA.mM$^{-1}$.cm$^{-2}$ (±0.03, n = 3) in 0.5 M KOH, a Limit of Detection (LOD) of ~0.17 mM, and a Limit of Quantification (LOQ) of ~0.58 mM. Density Functional Theory (DFT) was performed to understand the doping prompting in the reduced bandgap. The fabricated electrode displayed a rapid response time of 2 s and demonstrated 95% stability, excellent reproducibility, and selectivity, as confirmed by tests with several interfering species. A comprehensive evaluation of the electrode's performance using human blood serum highlighted its robustness and reliability for cholesterol detection in clinical settings, making it a promising tool for clinical and pharmaceutical applications.

**Graphical Abstract:**

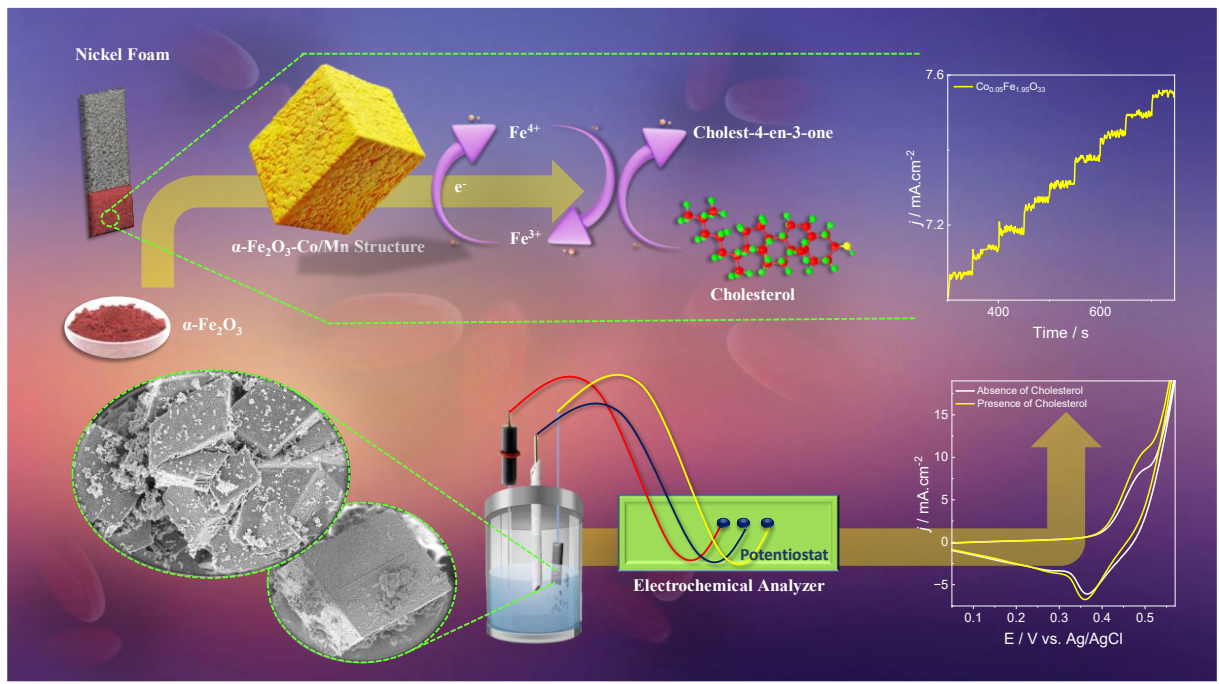

## 1. Introduction

Cardiovascular diseases (CVDs) are the leading cause of mortality worldwide, with growing public awareness since the 1980s regarding the risks associated with elevated blood cholesterol levels [1]. Cholesterol, a unique waxy lipid, is essential for various bodily functions, necessitating close monitoring [2]. Numerous studies have proven that elevated cholesterol noticeably raises the risk of CVDs [3]. In healthy individuals, cholesterol levels are below 200 mg/dL (5.17 mM), Levels around 240 mg/dL (6.21 mM) are associated with peripheral vascular disease, insulin-dependent diabetes, hypertension, and cardiovascular conditions [3–6].

Regular monitoring of cholesterol levels has become essential in contemporary healthcare, prompting the development of various methods to assess cholesterol concentrations accurately. Consequently, numerous techniques have emerged for detecting cholesterol in biological and food samples. These include spectrophotometric methods [7], electrochemical approaches [4], high-performance liquid chromatography (HPLC) [8], and enzymatic colorimetric techniques [9]. Although these techniques demonstrate significant effectiveness in cholesterol sensing, they still possess certain limitations, such as being time-consuming, high-cost, requiring expensive equipment, and potentially yielding background noise [10]. To tackle the problems based on these techniques, non-enzymatic biosensors that display strong thermal and chemical stability and good sensitivity and selectivity are showing a greater interest among researchers [11]. To bridge the gap, integrating non-enzymatic biosensors and nanotechnology has resulted in the creation of nanocomposites that considerably enhance the electrochemical performance of cholesterol sensors [12,13]. These advancements offer several predominant benefits, including improved selectivity, reduced cost, ease of handling, and rapid response times, rendering them highly effective for point-of-care testing [14]

The functionality of biosensors hinges on the selection and development of nanostructured materials [15] To develop electrochemically active substances for non-enzymatic sensing applications, various nanomaterials have been evaluated, including zero-dimensional (0D) nanomaterials [16], one-dimensional (1D) nanowires [17], and tubes [18], two-dimensional (2D) metal/metal oxide [19,20], graphenes [21], and three-dimensional (3D) nanoflowers [22,23], cubes [24], rods [25,26] and spheres [27].

Although advanced materials remain the primary focus of research, there is increasing interest in simpler nanocomposites that demonstrate comparable sensitivity and electrochemical performance. In recent years, transition metal oxides, including ZnO, CuO, $SnO_2$, $Fe_2O_3$, $Ag_2O$, $WO_3$, $NiO_2$, and $V_2O_5$, have been extensively investigated for their potential in non-enzymatic sensing applications [6,28–30]. These materials exhibit high efficacy owing to their metal centers and greater surface areas, thereby enhancing the density of electrochemically active sites in alkaline environments [16,31–36]. In the past few years, doping metal oxides has been shown to boost electrochemical performance and photocatalytic activity and improve electrical and photoelectrochemical features by boosting charge carrier density and conductivity [37].

Currently, activated carbon derived from Piper nigrum is being synthesized and combined with α-$Fe_2O_3$, followed by modification with a carbon paste electrode (APC-$Fe_2O_3$/CPE) [38] exhibited high sensitivity with a linear range of 25 nM to 300 nM, LOD of 8 nM, and LOQ of 26 nM. Additionally, electrolyte-gated transistor-based biosensors utilizing α-$Fe_2O_3$ decorated with ZnO nanorods demonstrated a broad linear range of 0.1 to 60 mM and a sensitivity of 37.34 $\mu A.mM^{-1}.cm^{-2}$ [39]. Moreover, the phase transition of $Fe_3O_4$ to α-$Fe_2O_3$ was achieved through electrophoretic film deposition on ITO-coated glass plates, which exhibited a sensitivity of 193 $nA.mg^{-1}$ dl $cm^{-2}$ with a linear range and response time of 25 –500 mg $dl^{-1}$ and 60 s, respectively [40]. Another study on the electrochemical behavior of citrate-modified β-cyclodextrin (CIT-BCD) and $Fe_3O_4$ synthesized via the coprecipitation method (CIT-BCD@$Fe_3O_4$) reported a linear range of 0 to 100 μM and a LOD of 3.93 μM [41].

If we come to a bimetallic nanocomposite glassy carbon electrode combined with $Cu_2O$/$MoS_2$ nanohybrid, which demonstrated the sensitivity of 111.74 $\mu A.\mu M^{-1}.cm^{-2}$ with LOD of 2.18 $\mu M$ and linear range spanning from 0.1–180 $\mu M$, showcased an effective alternative for cholesterol sensing[42]. Additionally, utilizing low-cost galvanic deposition, a ZnO/$WO_3$ composite demonstrates sensitivity of 176.6 $\mu A.cm^{-2}.mM^{-1}$ with a linear range of 0 – 320 μM, LOD of 5.5 nM[43]. Another study on the electrochemical study of NiO/CuO nanocomposite synthesized by the electrospinning method reported a sensitivity of 10.27 $\mu A.mM^{-1}.cm^{-2}$ linear range of 0.8 to 6.5 mM and a LOD of 5.9 μM[44]. Despite these advancements, challenges such as cost-effectiveness, lifespan, stability, and sensitivity to pH and temperature variations complicate the use of complex molecular sensors. To address these issues, our study aims to develop a transition metal-doped metal oxide cholesterol sensor to enhance selectivity, sensitivity, and stability while reducing costs. This article clearly shows the strategic comparative synthesis of bimetallic doping in $M_xFe_{2-x}O_3$ (M = Mn–Co), so far to the best of our knowledge, is the first comparative study investigating a nonenzymatic cholesterol sensor.

The primary goal of this work was to synthesize α-$Fe_2O_3$ through the hydrothermal approach and subsequently implement doping $M_xFe_{2-x}O_3$ (M = Mn - Co, and x = 0.01, 0.05, and 0.1). The doping concentrations of 1%, 5%, and 10% were selected to systematically investigate the variations in electronic structure, redox activity, and conductivity induced by doping. The obtained materials were fabricated on NF, and the electrodes were evaluated for sensitivity, LOD, LOQ, response time, and linear range. The comparison of the doped materials was intended to choose the most promising dopants, which were then subjected to a comparison analysis of their properties for biosensing applications. Electrochemical characteristics were assessed through Cyclic Voltammetry (CV), Chronoamperometry (CA), Differential Pulse Voltammetry (DPV), Electrochemical Impedance Spectroscopy (EIS), and the evaluation of electrochemical active surface area (ECSA). The selected material was further analyzed for its application in blood serum cholesterol detection.

## 2. Experimental Details
### 2.1 Chemicals and Materials

Nickel Foam ((NF), Thickness 0.5mm, 99.9% Purity) purchased from Global Nanotech Mumbai. Iron (III) Chloride Tetrahydrate ($FeCl_3 \cdot 4H_2O$, 99% Purity), Dopamine Hydrochloride (DA), Poly (Vinylidene Fluoride) (PVDF), Triton$^{TM}$ X – 100, and Urea ($CO(NH_2)_2$, 98% Purity) was procured through Sigma-Aldrich, Germany. Manganese (II) Acetate Tetrahydrate (($CH_3COO)_2Mn \cdot 4H_2O$, 98.5% Extra Pure), Cobalt (II) Acetate Tetrahydrate (($CH_3COO)_2Co \cdot 4H_2O$, 98.5% Extra Pure), Potassium Hydroxide Pellets (KOH, 85% Extra Pure), Uric Acid ((UA), $C_5H_4N_4O_3$, 99% AR), Cholesterol ($C_{27}H_{46}O$, 97% Extra Pure), L-Ascorbic Acid ((AA), $C_6H_8O_6$, 99% Extra Pure), N-Methyl 2-Pyrrolidone (NMP) ($C_5H_9NO$, 98% Purity), Potassium Chloride (KCl, 99% Purity), and Sodium Chloride (NaCl, 99.5% Purity) were purchased from LOBA Chemie Pvt. Ltd. Ultra-pure Milli-Q water (Elga Veolia) was utilized throughout the experiment. All chemicals were analytically certified and used without filtering.

## 2.2 Synthesis of $M_xFe_{2-x}O_3$ (M = Mn - Co and x = 0.01, 0.05, and 0.1)

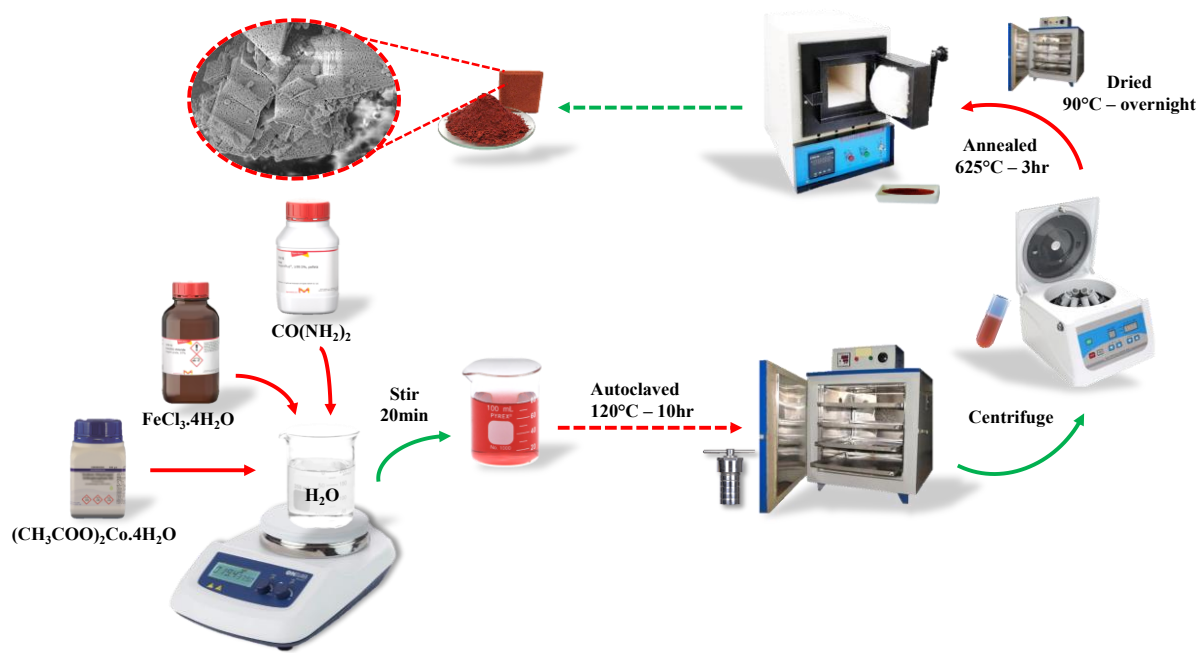

*Figure 1:* Schematic representation of the synthesis of $M_xFe_{2-x}O_3$ (M = Mn - Co, and x = 0.01, 0.05, and 0.1).

α-$Fe_2O_3$ and $M_xFe_{2-x}O_3$ nanostructures were synthesized by preparing a homogeneous solution of 0.1 M $FeCl_3.4H_2O$ and 0.1 M $CO(NH_2)_2$ dissolved in 80 mL of Milli-Q water for 30 minutes. To this solution, calculated quantities of Mn and Co precursors (with x = 0.01, 0.05, and 0.1) were added. The resulting mixture was loaded into a 100 mL autoclave and underwent hydrothermal processing at 120 °C for 10 hours. Once the reaction was completed, the mixture was allowed to reach room temperature, and collected by centrifugation. The precipitate was washed twice with Milli-Q water, followed by a single wash with ethanol. Once washed, the products were dried in an oven at 90 °C overnight and thereafter annealed at 625 °C for 3 hours. A schematic representation of the synthesis process for $M_xFe_{2-x}O_3$ is provided in **Fig. 1**.

The resulting products were designated with specific names based on their composition as $Co_{0.01}Fe_{1.99}O_3$ (CF1), $Co_{0.05}Fe_{1.95}O_3$ (CF5), $Co_{0.1}Fe_{1.90}O_3$ (CF10), $Mn_{0.01}Fe_{1.99}O_3$ (MF1), $Mn_{0.05}Fe_{1.95}O_3$ (MF5), and $Mn_{0.1}Fe_{1.90}O_3$ (MF10).

## 2.3 Fabrication of electrode

The synthesized nanocomposite comprising α-$Fe_2O_3$ and $M_xFe_{2-x}O_3$ (M = Mn - Co, and x = 0.01, 0.05, and 0.1) was fabricated on the NF electrode. Initially, accurately measured product samples were blended with PVDF in a 9:1 proportion in a mortar pestle. Subsequently, NMP was gradually added drop by drop till the mixture achieved paste consistency. The resultant paste was uniformly coated on a 1×1 cm$^2$ area of priorly treated NF and dried in a vacuum oven for an extended period of 24 hours at 60 °C, and the quantity of product loaded will be 10 ± 0.2 mg.

**2.4 Material and Electrochemical Characterisations.**

The structural properties that enabled the identification and characterization of distinct phases and crystalline compositions of the material under research were analyzed using a Rigaku Miniflex 600 powder X-ray diffraction (XRD) device. The measurement covers a range of 5° to 90° and a scanning rate of 3°per minute with monochromatic Cu-Kα radiation of wavelength 0.154 nm. The crystallite size (D) was calculated based on the Debye-Scherrer equation. (**Eq. 1**) [45].

$$D = \frac{K\lambda}{\beta \cos\theta} \qquad (1)$$

Here, K is the Scherrer constant, λ is the Cu-Kα radiation wavelength, β is the full width of half maximum (FWHM) of the peak, and θ is the Bragg angle. The phonon vibration modes were examined with a Confocal Raman Microscope integrated with a Compact Raman Spectrometer (Renishaw, UK) with an objective lens magnification of x50. The optical absorption spectra were examined with an ultraviolet-near infrared spectrophotometer (UV-vis-NIR, Lambda 950, Perkin Elmer, Singapore). The chemical composition of the synthesized sample was examined with the Thermo Fisher Scientific ESCALAB Xi+ X-ray Photoelectron Spectrophotometer (XPS) with Al Kα X-ray source (1,486.7 eV) for the analysis. Morphological imaging was performed using a field emission scanning electron microscope (FESEM) (7610FPLUS, Jeol, Japan) equipped with energy dispersive spectroscopy (EDAX) capabilities, allowing for a thorough investigation of the morphological characteristics.

The electrochemical characterization was conducted using an Autolab PGSTAT204 electrochemical workstation. CV experiments employed a three-electrode configuration, with the synthesized material coated on NF serving as the working electrode (WE), covering an area of 1×1 cm$^2$. A saturated Ag/AgCl electrode (with potassium chloride) served as the reference electrode (RE), while a platinum electrode functioned as the counter electrode (CE), both operating within a 0.5 M KOH electrolyte solution. CV analysis was carried out within a potential window of 0 to 0.75 V, with scan rates ranging from 5 to 120 mV/s. CA studies involved sequentially adding a cholesterol solution into the electrolyte under continuous stirring, maintained at a constant applied potential of +0.55 V vs. Ag/AgCl. DPV was executed over a potential range of 0.35 to 0.65 V vs. Ag/AgCl at a scan rate of 10 mV/s and a pulse amplitude of 50 mV. The ECSA was evaluated by varying the scan rates from 5 to 75 mV/s within a potential range. The double-layer capacitance ($C_{dl}$) was determined by plotting $\Delta j$ (where $\Delta j = j_a - j_c$) against the scan rate. The accurate surface area was then calculated using the formula $C_{dl}/C_s$ [46], where $C_s$ is the capacitance of an atomically smooth surface, taken as 40 $\mu F.cm^{-2}$ [47,48]. The sensitivity was evaluated from the calibration curve obtained by plotting current versus concentration, as expressed by the corresponding equation (**Eq. 2**).

$$\text{Sensitivity} = \frac{\text{Slope of the calibration curve}}{\text{Surface area}} \quad (2)$$

The formula used to calculate LOD and LOQ are represented by (**Eq. 3 and 4**);

$$\text{LOD} = \frac{3 \; X \; \text{Standard deviation}}{\text{Slope of the calibration curve}} \quad (3)$$

$$\text{LOQ} = \frac{10 \; X \; \text{Standard deviation}}{\text{Slope of the calibration curve}} \quad (4)$$

In order to ensure reproducibility, each test was carried out three times (n = 3), and the mean ± standard deviation is employed to depict the results. Statistical significance was regarded as variations with a $p < 0.03$ at the level of confidence set at 97%.

3. **Results and discussion**

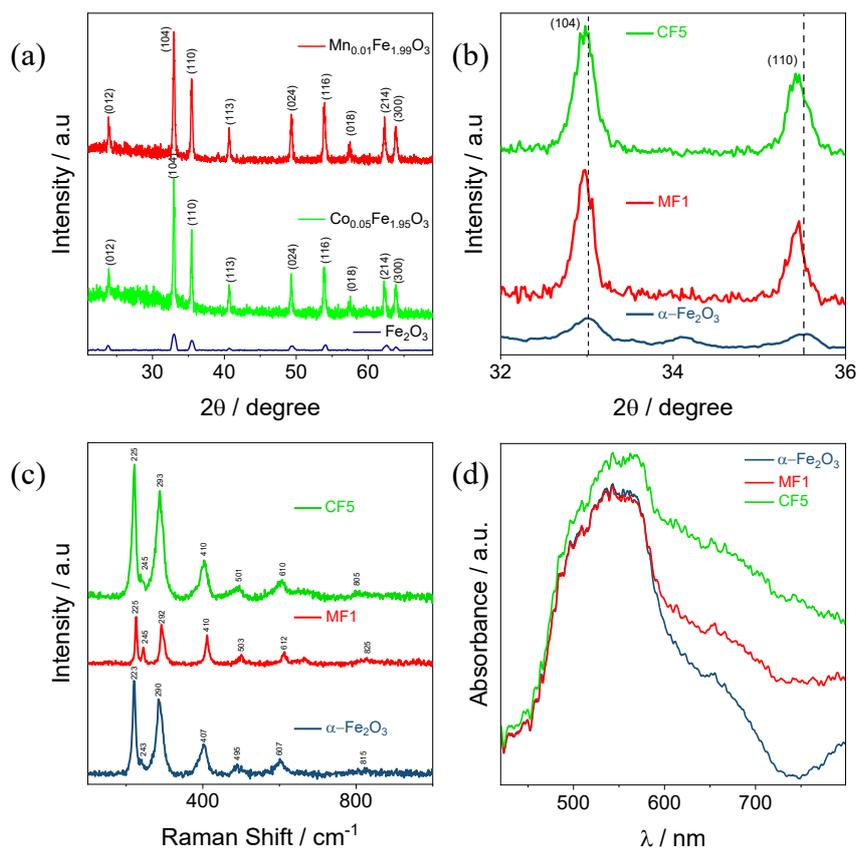

***Figure 2:*** *(a) XRD Spectra of α-Fe$_2$O$_3$, CF5, and MF1; (b) shift of the (104) and (110) XRD peaks to lower diffraction angles due to incorporation of substituent; (c) Raman spectra analysis of α-Fe$_2$O$_3$, MF1, and CF5 respectively; (d) UV-Vis profiles of α-Fe$_2$O$_3$, CF5, and MF1 respectively;*

The phase purity and composition of α-Fe$_2$O$_3$ and M$_x$Fe$_{2-x}$O$_3$ (M = Mn - Co, and x = 0.01, 0.05, and 0.1) were analyzed using powder XRD. **Fig. S1(a-c)** presents the XRD patterns of α-Fe$_2$O$_3$, and M$_x$Fe$_{2-x}$O$_3$, revealing a clear match with the JCPDS standard (No. 33-0664) and confirming its rhombohedral hexagonal phase. The non-appearance of additional impurity diffraction peaks in the XRD spectra confirms the pristine nature of α-Fe$_2$O$_3$. The diffraction peaks at 2θ angles of 23.83°, 32.99°, 35.46°, 40.69°, 49.47°, 54.08°, 57.14°, 62.57°, and 63.87° correspond to the (012), (104), (110), (113), (024), (116), (018), (214), and (300) planes of α-Fe$_2$O$_3$,[49].

**Fig. 2(a,b)** presents the XRD spectra of α-Fe$_2$O$_3$, CF5, and MF1 materials under investigation. A notable shift in the diffraction peaks toward lower angles, along with changes in peak intensity, confirms successful doping in the respective host structures. This shift is ascribed to lattice distortion induced by the incorporation of Co$^{3+}$ (0.61 Å) and Mn$^{3+}$ (0.645 Å), which possess smaller ionic radii compared to Fe$^{3+}$ (0.645 Å), into the α-Fe$_2$O$_3$ lattice. The crystallite

size, estimated through the Debye-Scherrer equation (**Eq. 1**) concerning the (104) diffraction plane, revealed an initial size of 16.18 nm for pure α-$Fe_2O_3$. In contrast, the doped samples MF1 and CF5 exhibited significantly larger crystallite sizes of 32.69 nm and 37.39 nm, correspondingly. This increase in crystallinity introduces defects in the α-$Fe_2O_3$ structure, enhancing the electron transfer kinetics, which accounts for the superior electrochemical performance of CF5/NF [50].

The Raman spectra analysis depicted in **Fig. 2(c)** provides valuable insights into the structural characteristics of nanoparticles for α-$Fe_2O_3$, MF1, and CF5. Additional Raman spectra data for the remaining dopants are presented in **Fig. S4(a,b)**. The vibrational spectra of pure α-$Fe_2O_3$ exhibit two distinct peaks at 223 and 495 $cm^{-1}$ corresponding to $A_{1g}$ modes, and five peaks at 243, 290, 407, 607, and 815 $cm^{-1}$ corresponding to $E_g$ modes, respectively. These seven peaks collectively indicate the characteristic vibrational modes of the α-$Fe_2O_3$ structure [49]. Notably, the absence of extra peaks signifies purity and confirms that α-$Fe_2O_3$ nanoparticles are produced without contaminants. The Raman spectra of MF1 and CF5 closely resemble those of pure α-$Fe_2O_3$, underscoring that the doped nanoparticles maintain their unique hematite structure. Despite the incorporation of Mn and Co dopants, the structural characteristics of α-$Fe_2O_3$ remain unchanged, as evidenced by the consistent presence of these features across the doped samples.

UV-vis spectra of α-$Fe_2O_3$, CF5, and MF1 were measured in terms of absorbance as illustrated in **Fig. 2(d)**. The band edge absorption for α-$Fe_2O_3$ typically occurs in the 520 - 565 nm wavelength range. In this study, the compound exhibited strong absorption at 542 nm, with CF5 and MF1 also demonstrating prominent peaks within this range. The band gap energies of the materials were evaluated through the Kubelka-Munk function $f(R)^2$ plotted against energy (eV) by utilizing reflectance spectra depicted in **Fig. S4(c)**. Linear extrapolation at $f(R)^2 = 0$ estimated the resultant band gap values: α-$Fe_2O_3$ (2.05 eV) > MF1 (2.02 eV) > CF5 (1.9 eV) **Fig. S4(d)**. The observed reduction in the bandgap can be attributed to the formation of additional energy levels proximate to the valence band edge, which reduces the energy requirement for electronic transitions from valence to conduction bands[51,52].

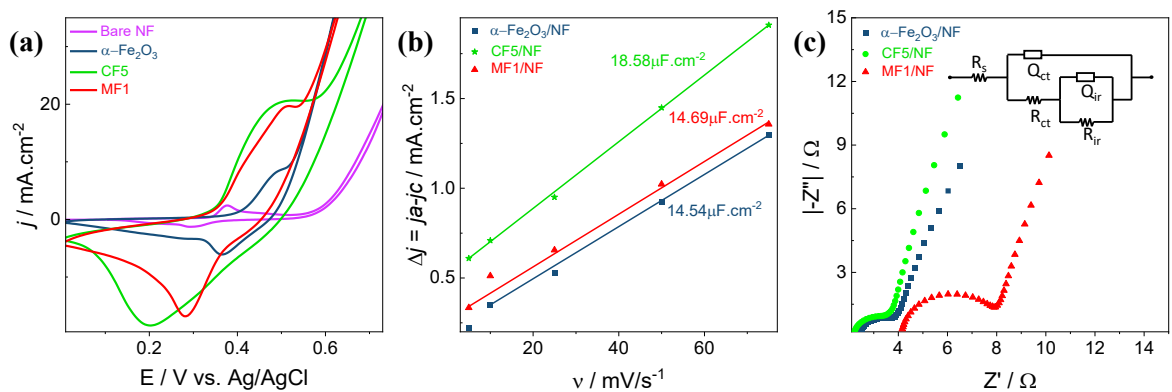

*Figure 3:* (a) Comparative analysis of CVs of Bare NF, α-Fe$_2$O$_3$/NF, CF5/NF, and MF1/NF in 0.5M KOH with scan rate of 50 mV/s; (b) ECSA Plot α-Fe$_2$O$_3$/NF, CF5/NF, and MF1/NF; (c) Nyquist Plot of α-Fe$_2$O$_3$/NF, CF5/NF, and MF1/NF respectively.

The electrocatalytic performance of bare NF, α-Fe$_2$O$_3$/NF, and M$_x$Fe$_{2-x}$O$_3$/NF (M = Mn - Co, and x = 0.01, 0.05, and 0.1) was evaluated by maintaining the scan rate of 50 mV/s over a potential window of 0 V to +0.75 V vs. Ag/AgCl using CV measurements in 0.5 M KOH as depicted in **Fig. S2(a-c)**. Among the doped materials, CF5/NF and MF1/NF demonstrated superior performance, which is demonstrated in **Fig. 3(a)**. Notably, among the three electrodes, CF5/NF proved superior electrocatalytic activity this is due to cobalt doping add more active sites to the α-Fe$_2$O$_3$ surface, making it easier for reactant molecules to adsorb and activate. Furthermore, cobalt ions can change the electronic structure of α-Fe$_2$O$_3$, resulting in improved charge transfer kinetics and catalytic activity[53]. This is evident from the anodic and cathodic peaks observed during redox reactions. To elucidate the surface-active sites linked to the number of electrons exchanged primarily in the oxidation process, a detailed study of the CV profiles of α-Fe$_2$O$_3$/NF, CF5/NF, and MF1/NF, are illustrated in **Fig. S3(a-c)**. The number of electron transfers for the three primary materials was determined to be $0.6 \times 10^{19}$ for α-Fe$_2$O$_3$/NF, $4.53 \times 10^{19}$ for MF1/NF, and $12.19 \times 10^{19}$ for CF5/NF [54,55]. The present study highlights that within three materials, CF5/NF exhibits the highest electron transfer rate during the oxidation process illustrating the better electrocatalytic performance of CF5/NF in alkaline electrolytes.

Unlike the BET method, the ECSA analysis entails the complete immersion of the material electrode in the electrolyte, enabling a precise assessment of its surface area activity, of all synthesised materials are depicted in **Fig. S5(a-g)** [48,54]. **Fig. 3(b)** reveals α-Fe$_2$O$_3$/NF, MF1/NF, and CF5/NF possessed 2C$_{dl}$ values of 14.54 μF.cm$^{-2}$, 14.69 μF.cm$^{-2}$, and 18.58 μF.cm$^{-2}$, accordingly. The ECSA results correlate with the genuine surface areas of each electrode as 0.182 cm$^2$, 0.184 cm$^2$, and 0.232 cm$^2$ in respective manner. Furthermore, a detailed evaluation of the surface area calibration plot and calculation are enclosed in **Fig. S5 (h, i)** and **Table S1**. The electrocatalytic performance results demonstrate that the CF5/NF electrode possesses the highest ECSA activity, indicating its superior potential for electrocatalytic applications due to an increased number of active sites.

To assess the efficiency of electrical charge transfer, EIS was performed, and the results were interpreted through Nyquist plots as presented in **Fig. 3(c)** to evaluate the resistance and capacitance of the material. For a more detailed interpretation, the Nyquist plots were modeled with the equivalent circuit $R_s(Q_{ct}(R_{ct}(Q_{ce}R_{ce})))$, with the fitted parameters provided in **Table S2**. The electrolyte resistance between the working and counter electrodes, represented by solution resistance ($R_s$), was determined from the x-intercept of the semicircle at high frequencies. Additionally, the electrode-electrolyte interface impedance, comprising parallel elements $Q_{ct}$ and $R_{ct}$, was included in the model. The internal resistance of the catalyst was represented by parallel elements $R_{ir}$ and $Q_{ir}$ in series with $R_{ct}$. Notably, the semicircle for CF5/NF displayed a smaller radius, indicating a reduction in internal and charge transfer impedance. This is also confirmed by the Mott-Schottky plot in **Fig. S4(e)**. This decrease in impedance and solution resistance correlates with the improved electrocatalytic performance observed for CF5/NF [56].

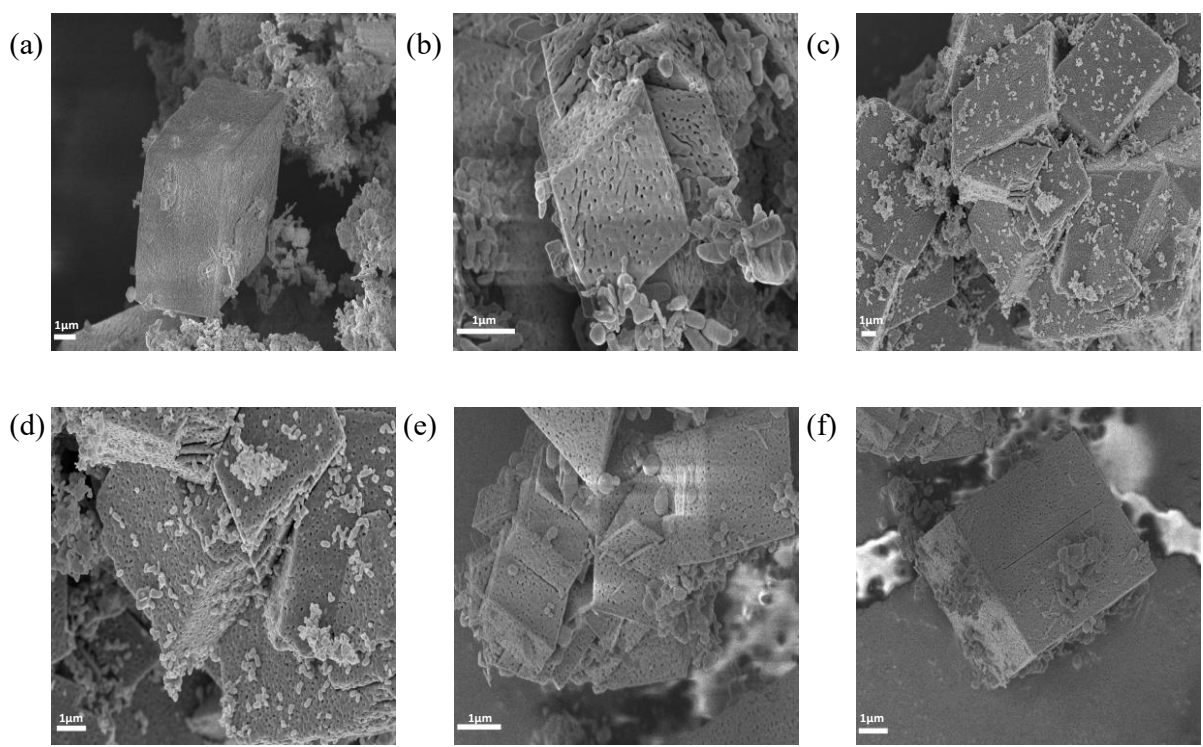

*Figure 4: FESEM images of (a,b) α-Fe$_2$O$_3$; (c, d) CF5; (e, f) MF1.*

The nanocomposites were morphologically characterized by FESEM and EDAX. FESEM images of α-Fe$_2$O$_3$, CF5, and MF1 nanoparticles are depicted in **Fig. 4(a-f),** respectively, illustrating their rhombus-like structure. The surface morphology is rough and porous, indicating the occurrence of α-Fe$_2$O$_3$ formation **Fig. 4(a,b)**. In CF5 and MF1, the incorporation of dopants into the lattice structure of α-Fe$_2$O$_3$ results in the formation of small granule-like structures on the surface of the framework; however, no surface defects are present within the lattice. Significantly, this observation implies that the α-Fe$_2$O$_3$ structure is not altered in CF5 and MF1.

**Fig. S6(c-e)** EDAX reveals the images showcasing a relatively uniform distribution of Fe and O components throughout the sample, confirming the homogeneity of the synthesized materials [57]. The low-intensity profiles observed for the Co and Mn dopants, in conjunction with the base elements, validate the effective incorporation of Co and Mn into the α-Fe$_2$O$_3$ matrix. Furthermore, EDAX analysis shows the existence of carbon (C) peaks at around 0.2 keV, which is attributed to the carbon tape used for sample handling during SEM and EDAX measurements.

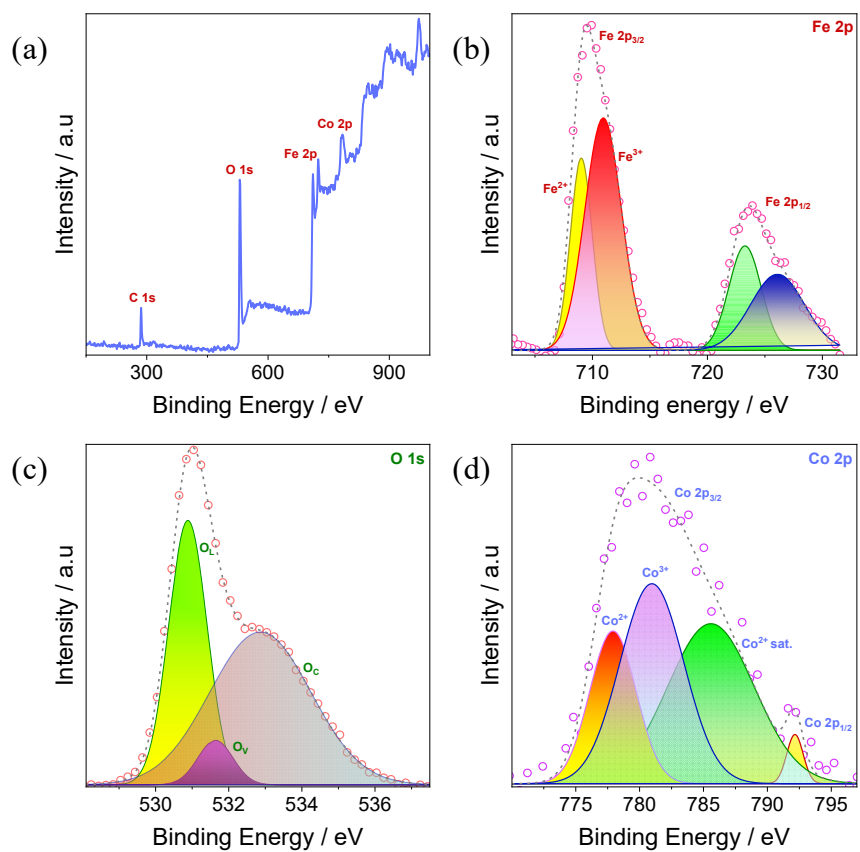

*Figure 5: XPS Spectrum of CF5 (a) Survey; (b) Fe 2p; (c) O 1s; (d) Co 2p.*

XPS was used to validate the surface chemical environment and oxidation state of CF5. Fe, O, Co, and C elements have been identified on the surface of the CF5, based on survey scans **Fig. 5a** carried out over an energy range of 100-1000 eV, with no observable traces of contaminants. Two distinguished peaks, individually splitting into doublets, were observed in the high-resolution Fe 2p XPS spectra **Fig. 5b**, demonstrating spin-orbit coupling in α-$Fe_2O_3$. In particular, peaks at 723.27 eV and 726.06 eV were found to correspond to $Fe^{2+}$ $2p_{1/2}$ and $Fe^{3+}$ $2p_{1/2}$, whereas peaks at 709 eV and 710.92 eV correlated with the binding energies of $Fe^{2+}$ $2p_{3/2}$ and $Fe^{3+}$ $2p_{3/2}$ respectively. The combined presence of $Fe^{2+}$ and $Fe^{3+}$ oxidation states points to a complex surface chemistry that may be impacted by dopant ion electron donation and oxygen vacancies from high-temperature annealing, among other factors. [58].

Well-defined peaks for the O 1s spectra of CF5 in **Fig. 5c** indicate the appearance of three distinct forms of surface oxygen species, which are represented by peaks at 530.87 eV involved with iron-oxygen bonding represented by the lowest energy peak, which is associated with lattice oxygen ($O_L$), 531.63 eV at the intermediate energy level, oxygen vacancies ($O_v$) appear as chemisorbed oxygen species, and 532.87 eV the final peak with the highest energy related to adsorbed oxygen species ($O_c$). The obtained results were consistent with prior literature findings [59,60].

Likewise, the Co 2p XPS spectrum in **Fig. 5d** shows an obvious peak at 780 eV, which implies the presence of Co in the CF5 lattice. This peak exhibits additional subdivision, with the Co $2p_{3/2}$ demonstrating three distinct peaks that show changes in the cobalt oxidation state. In particular, the $Co^{2+}$ peak appears at around 777.88 eV, while the $Co^{3+}$ peak is detected at 780.94 eV. Furthermore, a satellite peak of $Co^{2+}$ is seen at 785.61 eV, which could be ascribed to plasmons, energy loss processes, or shake-ups. In addition, the cobalt $2p_{1/2}$ peak has been observed at 792.16 eV, which adds to our understanding of the entire span of cobalt's oxidation states in the CF5 lattice [61,62].

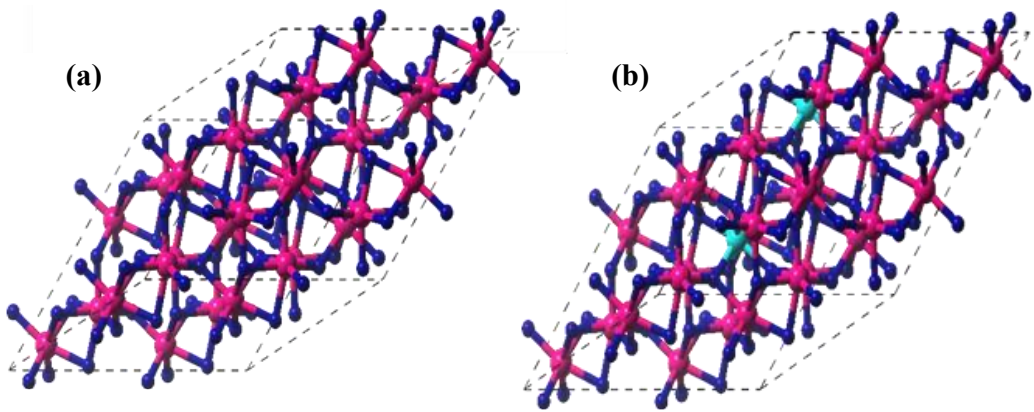

*Figure 6: Optimized geometry of (a) Fe₂O₃ and (b) CF5*

In this study, the Vienna Ab initio Simulation Package (VASP) was used to perform first-principles DFT simulations in order to examine the electronic structure and geometrical optimizations of both pristine and CF5 structures [63–65]. Exchange-correlation interactions were described using the Perdew-Burke-Ernzerhof (PBE) functional in the generalized gradient approximation (GGA), while electron-ion interactions were correctly represented using the projector augmented wave (PAW) approach [66,67]. To maintain a compromise between accuracy and computing economy, a plane-wave basis set with a kinetic energy cutoff of 520 eV was taken into consideration. For self-consistent computations, a Γ-centered 3×3×1 k-point mesh was used to sample the Brillouin zone. Using a U value of 5.0 eV based on the Dudarev and Botton technique [68], the Hubbard U correction was implemented within the GGA+U framework to account for strong electron correlations due to the presence of localized d-electrons in Fe and Co atoms. To minimize the force on each atom to less than 0.01 eV/Å, structural optimizations were performed. The energy convergence criterion between subsequent self-consistent stages was set at $1\times10^{-5}$ eV. After building the bulk haematite structure to analyze its structural and electrical characteristics, two Fe atoms were swapped out for Co atoms in a supercell structure consisting of 80 atoms to study the effects of Co doping. A thorough examination of the structural and electrical properties of both pristine and CF5 structures was made possible by these computational methods.

Structural properties:

DFT computations were used to examine the structural characteristics of pristine α-Fe$_2$O$_3$ and CF5, as shown in **Fig. 6(a, b)**. Fe atoms occupy octahedral positions inside a hexagonal close-packed lattice of oxygen (O) atoms in α-hematite's rhombohedral structure (R-3c). For pristine α-Fe$_2$O$_3$, the optimized lattice parameters were found to be a = 5.44 Å and α = 55.270°, which are values that closely match previous theoretical research and experimental results. The haematite structure's intrinsic cation-anion interactions are reflected in the Fe-O bond lengths within the octahedral units, which vary from 1.96 Å to 2.10 Å.

Minimal structural changes are brought about by the addition of 5% cobalt substitution, in which $Co^{3+}$ ions take the place of $Fe^{3+}$ ions. Optimal values of a = 5.42 Å and α = 55.270° are obtained by a modest contraction of the lattice parameters because of the lesser ionic radius of $Co^{3+}$ (roughly 0.61 Å) compared to $Fe^{3+}$ (about 0.645 Å). The unit cell volume decreases as a result of this little drop in lattice characteristics, indicating that cobalt was successfully incorporated into the haematite lattice without causing significant structural deformities.

Additionally, there are minor variations in the bond lengths between the Co-substituted Fe-O octahedra and pure Fe-O bonds. The Co-O distances, which range from 1.95 Å to 1.97 Å, are marginally shorter than the comparable Fe-O bonds. The stronger Co-O interactions brought about by the more localized electronic character of the Co 3d states are responsible for this change.

Electronic Properties of α-Fe$_2$O$_3$ and CF5:

The electronic structure of a sensing material plays a key role in defining its conductivity, charge transfer efficiency, and surface reactivity, all of which directly influence its sensing ability. In biosensing applications, α-Fe$_2$O$_3$ is a promising material because of its surface activity, tunable bandgap [69–71], and a stable semiconducting nature. However, its efficiency is limited by its weak electrical conductivity and comparatively high bandgap. In order to alter the electrical structure and possibly lower the bandgap, introduce mid-gap states, and improve charge carrier mobility, Co doping is investigated.

Co-incorporation's effects on the material's electronic conductivity and surface interaction potential can be understood by examining the density of states (DOS) and band structure of both pure and CF5. Since enhanced conductivity and charge transfer capacities can improve the sensor's responsiveness, this information is crucial for optimizing hematite-based materials for cholesterol detection. The creation of more effective biosensors is guided by the fundamental insights into the material's potential as an active sensing element that the electronic structure analysis offers.

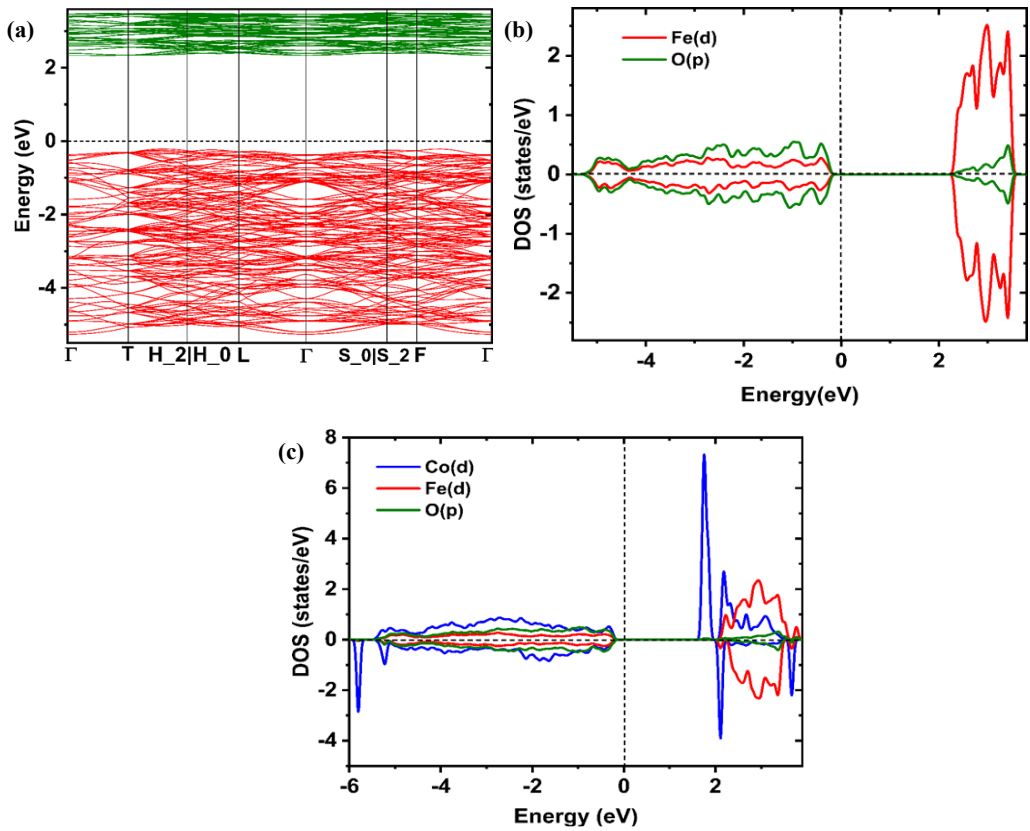

*Figure 7:* (a) Electronic band structure of α-$Fe_2O_3$ in the Brillouin zone of the rhombohedral cell, emphasizing the band gap along the high-symmetry path $\Gamma$-$T$-$H_2$|$H_0$-$L$-$\Gamma$-$S_0$|$S_2$-$F$-$\Gamma$. (b) α-$Fe_2O_3$ density of states (DOS), showing the band gap and the distribution of electronic states (c) The electronic density of states (DOS) of CF5 demonstrates the effect of Co doping on electronic characteristics and band gaps.

Band Structure Analysis:

Pristine α-$Fe_2O_3$ is an indirect bandgap semiconductor, according to band structure calculations, with the conduction band minimum (CBM) at a separate k-point and the valence band maximum (VBM) at the Γ-point. The calculated bandgap, which is shown in **Fig. 7(a)**, is 2.3 eV, which is the same value as was previously published. Fe 3d orbitals predominate in the conduction band, whilst O 2p orbitals contribute to the majority of the valence band. The electrical structure is significantly influenced by the Fe 3d-O 2p hybridization.

The band structure changes noticeably when 5% Co is doped at Fe sites. The bandgap drops to 1.93 eV, suggesting the formation of impurity states close to the conduction band as shown in **Fig. 7(c)**. Co 3d orbitals, which add more electronic channels and may improve electrical conductivity, are the source of these states. Moreover, a shift in the Fe 3d bands results from a minor perturbation of the Fe electronic states caused by Co inclusion.

Moreover, compared to normal LDA/GGA, which frequently underestimates the band gap, the LDA+U technique tends to increase it [72,73]. Compared to the conventional DFT approaches, the use of U helps to correct the electron-electron interactions, especially in the Fe 3d orbitals, which results in a bigger bandgap and a more accurate representation of the electronic structure [74].

DOS Analysis:

As seen in **Fig. 7(b,c)**, the atom projected density of states (PDOS) offers vital information on the electrical structure of both pure and CF5. O 2p states contribute to the majority of the valence band in pure α-$Fe_2O_3$, with Fe 3d states contributing close to the VBM. Fe 3d orbitals contribute to the conduction band, which has a wide band gap and restricts electrical conductivity. The DOS undergoes notable changes at 5% Co doping, especially in the vicinity of the CBM. By adding Co 3d states, more electronic states are produced close to the Fermi level, which lowers the bandgap and increases the concentration of carriers. This change suggests increased Fe-Co hybridization-induced electrical conductivity. Additionally, the doped system is better suited for applications needing effective charge transfer because the presence of Co states close to the CBM promotes electron transport.

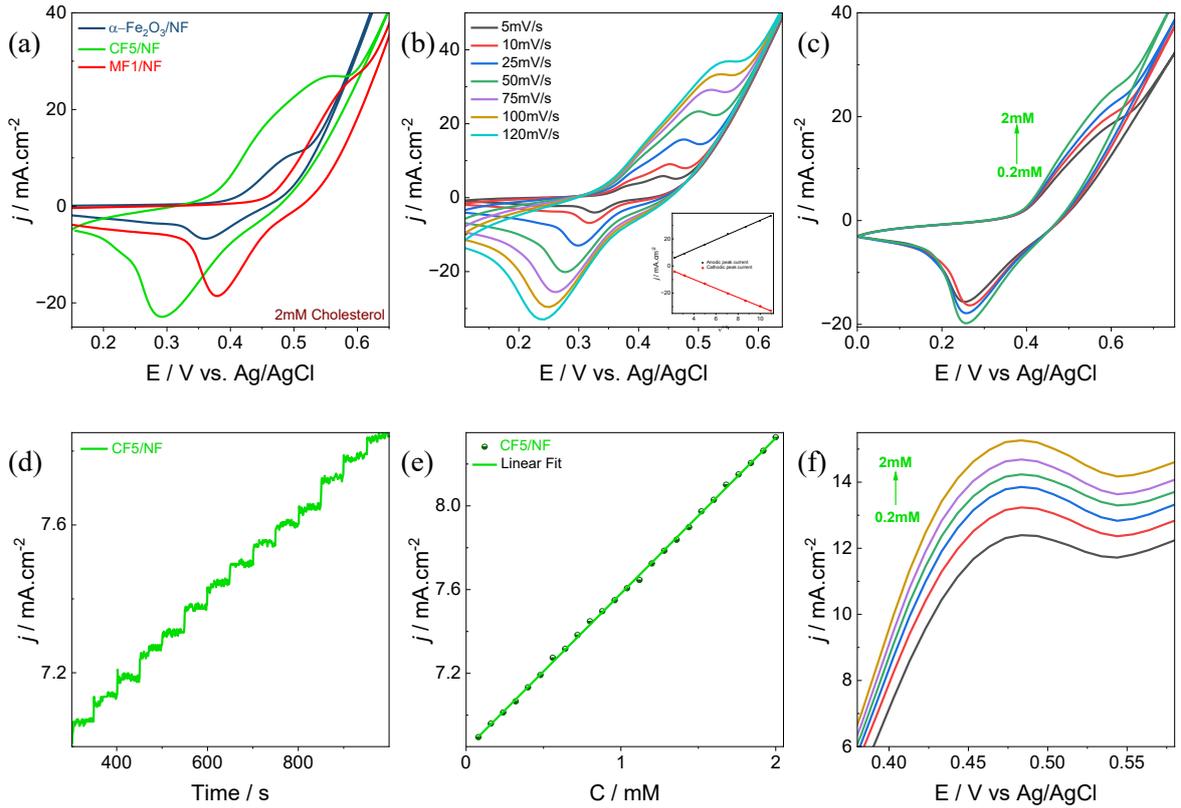

*Figure 8:* *(a) Comparison of CVs in the influence of 2mM Cholesterol in 0.5 M KOH with the scan rate of 50 mV/s for α-Fe$_2$O$_3$/NF, CF5/NF, and MF1/NF; (b) Scan rate study of CF5/NF in 0.5 M KOH from 5-120 mV/s. The anodic and cathodic peak currents are inserted as a function of the square root of the scan rate (inset); (c) The successive addition analysis for CF5/NF at the scan rate of 50 mV/s; (d) The CA study of CF5/NF was conducted in a 0.5 M KOH solution, the measurements were performed under stirring conditions, with an applied potential of +0.55 V; (e) Calibration plot of CF5/NF to estimate cholesterol; (f) DPV study of CF5/NF to analyze the electrochemical response to successive cholesterol additions.*

The CV analysis showed a progressive enhancement in both oxidation and reduction peak currents upon exposure to 2 mM cholesterol for α-Fe$_2$O$_3$/NF, CF5/NF, and MF1/NF in 0.5 M KOH electrolyte, at a scan rate of 50 mV/s, depicted in **Fig. 8(a)**. In particular, the CF5/NF CV analysis with and without the addition of cholesterol at the scan rate of 50 mV/s is determined in **Fig. S8(a)**, which reveals the efficiency of CF5/NF towards cholesterol sensing. The improved cholesterol-sensing performance of CF5/NF is ascribed to its well-defined crystalline structure, enhanced surface area, and superior electron transfer capabilities. Additionally, modifications in band gap energy, as indicated by the DFT study, contribute to a significant enhancement in electrocatalytic activity toward cholesterol oxidation. These factors collectively underscore the potential of CF5/NF for advanced cholesterol biosensing applications. **Eqs (5), (6)**, and **(7)** represent the detailed mechanism [75].

$$Fe_2O_3 (III) + OH^- \rightarrow Fe_2O_3 (III) - O_{ads} \quad (5)$$

$$Fe_2O_3 (III) - O_{ads} \rightarrow Fe_2O_3 (IV) = O + e^- + H^+ \quad (6)$$

$$Fe_2O_3 (IV) = O + Cholesterol \rightarrow Cholest\text{-}4\text{-}en\text{-}3\text{-}one + Fe_2O_3 (III) - O_{ads} + e^- + H^+ \quad (7)$$

α-$Fe_2O_3$ contributes structural stability and a porous surface with an extensive surface area, making it an ideal host matrix. The mechanism can be illustrated as follows: Initially, as presented in **Eq. 5**, the hydroxide ion (OH⁻) present in the electrolyte adsorbs onto the surface of $Fe_2O_3$ leading to the generation of adsorbed oxygen species ($O_{ads}$) on $Fe_2O_3$ (III). These adsorbed oxygen species undergo oxidation, leading to the formation of $Fe_2O_3$ (IV) with a higher oxidation state, along with the release of a proton (H⁺) and an electron (e⁻). Upon the addition of cholesterol to the electrolyte, $Fe_2O_3$ (IV) reacts, facilitates the formation of cholest-4-en-3-one. During this interval, $Fe_2O_3$ (IV) is reduced back to $Fe_2O_3$ (III) with an adsorbed oxygen species, releasing an additional e⁻ and H⁺. It was noted the release of two electrons during this process, aligns with the results obtained from the Laviron equation **Eq. 8**.

$$\text{Slope} = \frac{RT}{F\alpha n} \quad (8)$$

**Fig. S7(a)** gives the implementation of the Laviron equation, obtained by graphing the slope of ln ($v$) against the oxidation peak potential (E). This equation involves the ideal gas constant (R = 8.314 J), Faraday's constant (F = 96,480 C mol), temperature (T = 300 K), transfer coefficient ($\alpha$ = 0.5), and number of electrons associated with cholesterol oxidation (n) [34]. The study showed that 1.72 electrons are exchanged during cholesterol oxidation, closely approximating a two-electron transfer mechanism, which is further corroborated by DFT analysis.

**Fig. 8(b)** illustrates the scan rate study of CF5/NF, encompassing a potential range from 5 mV/s to 120 mV/s, on the other hand, **Fig. S8(b)** depicts the scan rate study of α-$Fe_2O_3$/NF. With an increase in scan rate, there is a corresponding increase in the peak potentials of both oxidation and reduction processes. Under reduced scan rates, the formation of a diffusion layer between the electroactive species and the electrode surface results in reduced interaction, leading to lower anodic and cathodic peak currents. Conversely, an increase in scan rate leads to the breakdown of the diffusion layer, and at a certain scan rate, it vanishes. This results in a shift of the anodic peak towards more positive potentials and the cathodic peak towards more negative potentials with increasing scan rates. Additionally, a linear correlation was observed between the square root of the scan rate and the redox peak currents of CF5/NF, as shown in the inset of **Fig. 8(b)** indicating the formation of the diffusion layer impedes the interaction between electroactive species and the electrode surface, thereby facilitating electron transfer through an outer-sphere mechanism.

The subsequent study, illustrated in **Fig. 8(c)** and **Fig. S8(c)**, presents the effects of successive additions of cholesterol within a linear concentration range of 0.2 mM to 2 mM in 0.5 M KOH, at a scan rate of 50 mV/s for CF5/NF, and α-$Fe_2O_3$/NF respectively. The data indicate a progressive rise in redox peak currents upon each incremental addition of cholesterol, signifying effective electrocatalytic oxidation and reduction of cholesterol. This observed enhancement in redox peak currents can be ascribed to the forming of a diffusion layer, facilitating improved interaction among the electroactive species and the electrode surface.

**Fig. 8(d)** depicts the CA current response of the CF5/NF electrode at an applied potential of +0.55 V in response to the sequential addition of cholesterol. Each addition of cholesterol results in a stepwise increase in current every 50 s, indicating the electrode's capacity to detect cholesterol effectively at each increment. Comparative CA responses for the α-$Fe_2O_3$/NF electrode are presented in **Fig. S8(d)**. At the specified anodic peak potential, the absence of a diffusion layer permits direct interaction between the electroactive species and the electrode surface, resulting in a stepwise increase in current upon each cholesterol addition. **Fig. 8(e)** presents the calibration curve, demonstrating the linear response across a cholesterol concentration range from 0.2 mM to 2 mM. The CF5/NF electrode exhibited high sensitivity, measured at 1364.2 µA.mM$^{-1}$.cm$^{-2}$ (±0.03, n = 3) as depicted in **Table S3**. Furthermore, the LOQ and LOD for the CF5/NF electrode were calculated to be ~0.58 mM and ~0.17 mM, respectively, with a response time of 2 sec **Fig. S9(a)**. In contrast, the α-$Fe_2O_3$/NF electrode demonstrated a sensitivity of 590 µA.mM$^{-1}$.cm$^{-2}$ (±0.05, n = 3) represented through **Fig. S8(e)** and **Table S3**.

Furthermore, DPV analysis was conducted to examine the electrochemical behavior of the CF5/NF electrode in the potential range of 0.35 to 0.65 V with a scan rate of 10 mV/s in 0.5 M KOH, as shown in **Fig. 8(f)**. A notable increase in peak currents corresponded to the redox reactions as the cholesterol concentration increased, confirming the electrode's catalytic activity. These findings underscore that CF5/NF effectively responds to cholesterol due to its crystal structure, high surface area, excellent electron transfer, decrease in charge transfer impedance, increased surface active sites, and due to defects of the CF5/NF electrode for cholesterol biosensing applications, demonstrating its potential for precise and dependable measurement of cholesterol levels.

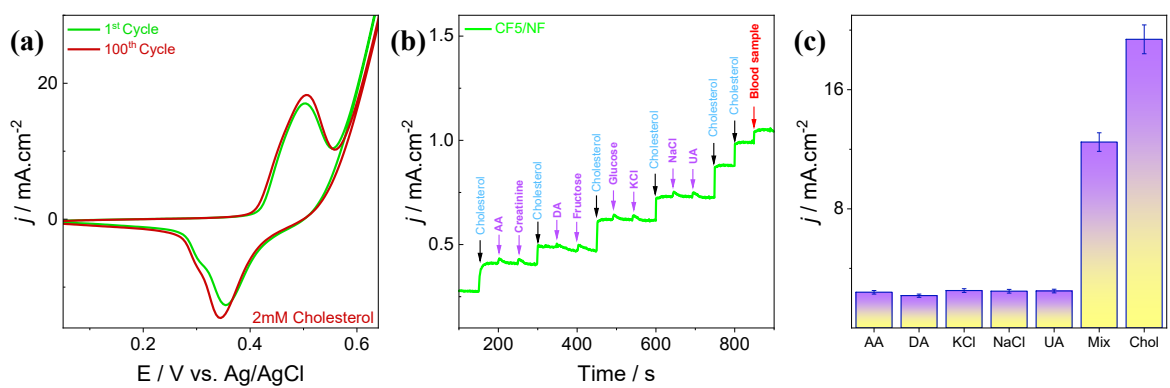

*Figure 9: (a) CVs of CF5/NF for 100 cycles in a 0.5 M KOH solution containing 2 mM cholesterol, with an applied potential scan rate of 50 mV/s; (b) An interference study of CF5/NF in a 0.5 M KOH solution under stirring conditions to assess its reaction to various interfering species of 1mM at a potential of +0.55 V; (c) DPV interference study of CF5/NF at a scan rate of 10 mV/s and pulse amplitude of 50 mV.*

To confirm the long-term reproducibility of the CF5/NF electrode, CV for 100 cycles in 0.5 M KOH containing 2 mM cholesterol was done. The resulting data, depicted in **Fig.9(a)**, exhibited minimal changes in oxidation and reduction peak potentials, retaining approximately 95% of its initial characteristics even after 100 cycles, while the α-$Fe_2O_3$/NF electrode retained 84.4% as presented in **Fig. S8(f)**. Furthermore, the reproducibility of the CF5/NF electrode was assessed by testing three different electrodes in a 0.5 M KOH solution, as shown in **Fig. S10(a)**. Where electrodes demonstrated consistent current ranges, thereby confirming their reproducibility. Additionally, the stability of the CF5/NF electrode was evaluated for 90 days depicted in **Fig. S10(b)** indicating that the electrode maintained stability with only a slight decrease in the current range, without any drastic changes. This indicates remarkable stability and reproducibility in the electrode's electrochemical performance. The study demonstrates that the CF5/NF electrode can maintain its catalytic activity and structural integrity through repeated use, underscoring its effectiveness for consistent and reliable cholesterol sensing.

One of the predominant characteristics of an effective biosensor is selectivity [76,77]. To evaluate the selectivity of the CF5/NF electrode as a biosensor, an interference analysis was conducted using CA under stirring conditions at a constant potential of +0.55 V in 0.5 M KOH. The electrode was examined with various interfering species, including ascorbic acid (AA), creatinine, dopamine (DA), fructose, glucose, KCl, NaCl, and uric acid (UA), with two different concentrations signifying 0.5 mM and 1 mM of interfering species as demonstrated in **Fig. S9(a)** and **Fig. 9(b)** respectively. The results indicated a stepwise increase in current exclusively upon adding cholesterol, with no significant current rise observed during the addition of the interfering species. This demonstrates that the CF5/NF electrode is highly selective for cholesterol biosensing. Moreover, an increase in current was observed upon adding a blood sample containing cholesterol, further confirming the electrode's sensitivity to cholesterol. To complement the interference study with CA, DPV analysis was also performed, as shown in **Fig. 9(c)**. This analysis included the individual interfering species, a mixture of interfering species with cholesterol, and cholesterol alone. The results showed no significant interference from the interfering species in the mixture, while there was a noticeable increase in the current for cholesterol. These findings confirm the CF5/NF electrode's excellent sensing properties for cholesterol, demonstrating high selectivity, sensitivity, and reliability for cholesterol biosensing applications.

**Table 1** demonstrates CF5/NF electrode biosensing capabilities to, those of existing metal oxide-based cholesterol biosensors reported in the literature.

| Electrode Material | Sensitivity ($\mu A \cdot mM^{-1} \cdot cm^{-2}$) | LOD (mM) | Linear range (mM) | Ref |
|---|---|---|---|---|
| ZnO nanorods | 4.2 | 1.78 | 1 - 9 | 78 |
| $Cu_2O$-$MoS_2$ | 73.55 | 0.0036 | 0.0001 – 0.18 | 42 |
| $NiO/MoS_2$ | 7.95 | 0.209 | 0.259 – 3.88 | 79 |
| $Fe_3O_4@SiO_2$/ MWNT | - | 0.005 | 0.01 - 4 | 80 |
| Oxidized Zn-In nanostructure. | 81 | - | 0.5 - 09 | 81 |
| Calcein-Co$_3$O$_4$ NCs | - | 0.00048 | 0.001 – 0.05 | 82 |
| PMO-BMCP | 226.30 | 0.00111 | 0.00002 – 333.3 | 83 |
| **CF5/NF** | **1364.2 (± 0.03, n = 3)** | **~ 0.17** | **0.2 - 2** | **This work** |

To further validate the performance of the CF5/NF electrode, we conducted real sample analysis using known cholesterol concentrations in human blood serum collected from nearby hospitals. Initially, the electrode was tested with synthetically available cholesterol twice at a concentration of 2 mM and then with a real sample as depicted in **Fig. S9(b)**. For the real sample analysis, the data obtained were compared with theoretical values provided by the hospitals. As summarized in the table below, the results showed an accuracy level of approximately 94 - 98%. This high degree of accuracy underscores the CF5/NF electrode's excellent sensing properties for cholesterol, confirming its potential for practical application in clinical diagnostics and real-world cholesterol analysis.

**Table 2** presents the analysis of human blood serum for cholesterol testing.

| Patient | Gender | Clinically Tested Result (mg/dL) | Experimental Result (mg/dL) | % Accuracy |
|---|---|---|---|---|
| **Patient 1** | Female | 308 | 291.6 | 94.68% |
| **Patient 2** | Female | 182 | 174.7 | 96% |
| **Patient 3** | Male | 211 | 201.2 | 95.36% |
| **Patient 4** | Male | 202 | 195.3 | 96.66% |
| **Patient 5** | Female | 200 | 196 | 98% |
| **Patient 6** | Female | 260 | 252.2 | 97% |

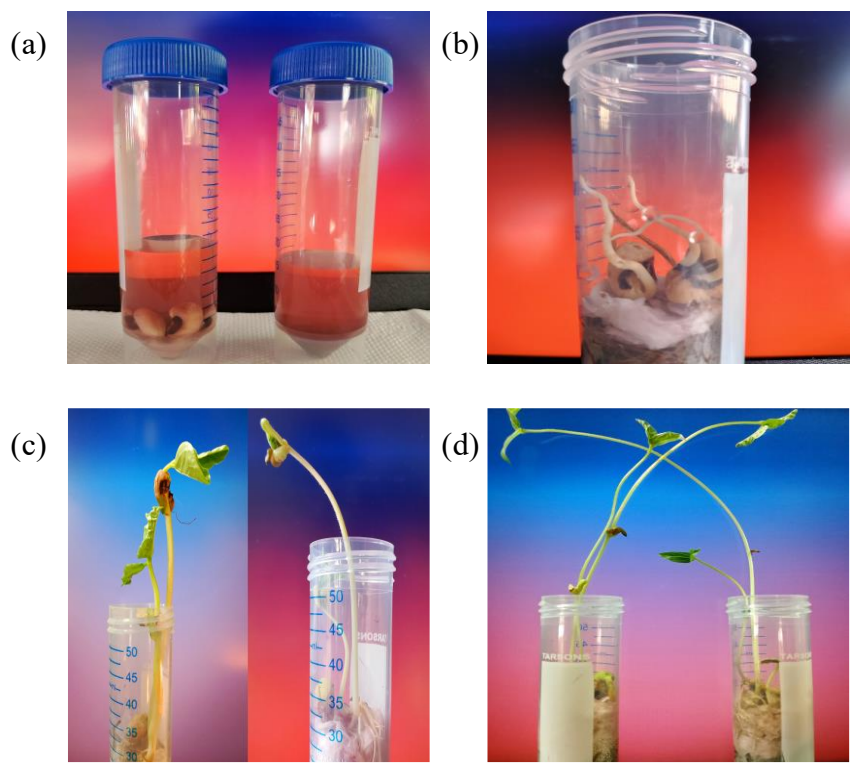

***Figure 10:*** *Preliminary study of biocompatibility demonstrating enhanced plant growth using CF5 solution, with photographs demonstrating seed germination on (a) Day 1; (b) Day 3; (c) Day 8; and (d) Day 13.*

Here in, a primary investigation towards non-toxic behavior and biocompatibility of a CF5 was encouraged by plant growth in addition to an electrochemical investigation. The CF5 solution is simply made by dissolving the synthesized product in water, highlighting its low environmental impact as shown in **Fig. 10(a-d)**. Green gram and horse gram seeds were selected as the subjects for the study. The seeds were initially immersed in the solution, and by day 3, germination was observed, as shown by the appearance of tiny buds. By day 13, healthy seedling growth was noticeable [84]. These outcomes underscore that the material CF5 does not exhibit any phytotoxicity and demonstrate the importance of examining the non-toxic behavior of nanomaterials, particularly when aimed at biosensing applications.

**Conclusions**

This study successfully synthesized $\alpha$-$Fe_2O_3$ and $M_xFe_{2-x}O_3$ (M = Mn - Co and x = 0.01, 0.05, and 0.1) and fabricated a working electrode by coating the materials on NF. Comprehensive structural, optical, morphological, and electrochemical analyses identified CF5/NF as a superior electrode material, exhibiting excellent sensitivity, a high surface area, and efficient electron transfer. The CF5/NF electrode demonstrated remarkable performance in cholesterol sensing, with a linear detection range from 0.2 mM to 2 mM, a sensitivity of 1364.2 $\mu A.mM^{-1}.cm^{-2}$ (± 0.03, n = 3) in 0.5 M KOH, a LOD of ~0.17 mM, and a LOQ of ~0.58 mM. Additionally, the electrode exhibited a rapid response time of 2 s. The evaluation of real samples revealed that the CF5/NF electrode exhibited outstanding cholesterol-sensing capabilities, demonstrating negligible interference from other species and underscoring its potential for enzyme-free cholesterol detection. The study concludes that the CF5 nanocomposite is a prominent nanocomposite for cholesterol sensing due to its robust electrochemical characteristics, non-toxic behavior, biocompatibility, and efficacy in detecting cholesterol in human blood samples, signifying a viable tool for clinical and pharmaceutical applications. This research shows good results over laboratory scale synthesis, expanding on this information's the future research is mainly focused on scaling up the synthetic procedure for a miniaturized and wearable platform for real-time point-of-care cholesterol sensing.

**CRediT authorship contribution statement**

**Conflict of interest**

The authors declare that they have no known competing financial interests or personal relationships that could have appeared to influence the work reported in this paper

**Acknowledgment**

The author (SS) expresses sincere gratitude to Endodiab Clinic, Mangalore, for the provision of blood serum samples and extends appreciation to the National Institute of Technology Karnataka, Surathkal, for their invaluable assistance in supporting this research and financial assistance for publication. Special thanks are also directed to the Central Research Facility (CRF), NITK for offering essential characterization facilities for the successful completion of this work. Authors KC and SR acknowledges the support by the Ministry of Education, Youth and Sports of the Czech Republic through the e-INFRA CZ (ID:90254), This research has been supported by the project QM4ST with. No.CZ.02.01.01/00/22_008/0004572.

**Ethical Approval**

The utilization of human serum samples in this research was authorized by the Ethical Committee of the National Institute of Technology Karnataka, Surathkal, under reference number NITK/Bioethics/2023/02, dated 24 April 2023. Experimental protocols were meticulously followed in accordance with institutional regulations and legal frameworks, ensuring ethical research conduct.